\documentclass[english] {sophia_rcs}

\title{Modeling Entanglement-Based Quantum Key Distribution for the NASA Quantum Communications Analysis Suite}

\author{Michael J. P. Kuban, Ian R. Nemitz, Yousef K. Chahine\\
	\normalsize National Aeronautics and Space Administration\\ Glenn Research Center\\ Cleveland, Ohio 44135\\
}

\begin{document}

\maketitle

\lhead{}
\chead{}
\rhead{}

\begin{abstract}
	One of the most practical, and sought after, applications of quantum mechanics in the field of information science is the use of entanglement distribution to communicate quantum information effectively. Similar to the continued improvements of functional quantum computers over the past decade, advances in demonstrations of entanglement distribution over long distances may enable new applications in aeronautics and space communications. The existing NASA Quantum Communications Analysis Suite (NQCAS) software models such applications, but limited experimental data exists to verify the model's theoretical results. There is, however, a large body of experimental data in the relevant literature for entanglement-based quantum key distribution (QKD). This paper details a Monte Carlo-based QKD model that uses NQCAS input parameters to generate an estimated QKD link budget for verification of NQCAS. The model generates link budget statistics like key rates, error rates, and S values that can then be compared to the experimental values in the literature. Preliminary comparisons show many similarities between the simulated and experimental data, supporting the model's validity. A verified NQCAS model will inform experimental work conducted in Glenn Research Center's (GRC) NASA Quantum Metrology Laboratory (NQML), supporting the United States Quantum Initiative and potential NASA missions.

\end{abstract}

\begin{nomenclature}
\\NQCAS- NASA Quantum Communications Analysis Suite\\
QKD- Quantum Key Distribution\\
NQML- NASA Quantum Metrology Laboratory\\
QISE- Quantum Information Science and Engineering\\
RF- Radio Frequency\\
FSO- Free Space Optical\\
SPDC- Spontaneous Parametric Down-Conversion\\
EPR- Einstein-Podolsky-Rosen\\
CHSH- Clauser-Horne-Shimony-Holt\\
QBER- Quantum Bit Error Rate\\
PNR- Photon Number Resolution\\
POVM- Positive Operator-Valued Measure\\
CW- Continuous-Wave\\

\end{nomenclature}

\begin{roadmap}
	This paper begins with a brief introduction to QKD and the model's justification in \cref{sec:introduction}\hspace{2.5mm} This is followed by a thorough description of the model's implementation in \cref{sec:modelImplementation}\hspace{2.5mm} Preliminary data gathered from the simulation and the associated analysis can be found in \cref{sec:results}\hspace{2.5mm} Concluding remarks and future work can be found in \cref{sec:conclusion}\hspace{2.5mm} Finally, specific calculations mentioned in the paper can be found in \cref{sec:appen}
\end{roadmap}

%~~~~~~~~~~~~~~~~~~~~~~~~~~~~~~~~~~~~~~~~
% Seções
%~~~~~~~~~~~~~~~~~~~~~~~~~~~~~~~~~~~~~~~~
\section{INTRODUCTION} \label{sec:introduction}

Section I is meant to provide a pedagogical overview of quantum key distribution (QKD), as well as justification for the associated NASA Quantum Communications Analysis Suite (NQCAS) verification model. It begins by describing the theory behind QKD and the reasons for its popularity, then it moves on to a description of a common entanglement-based QKD protocol known as E91. Finally, it concludes with a description of why modeling this process is important for both NQCAS and for NASA as a whole.

\subsection{QKD Background}

The field of quantum information science and engineering (QISE) has become a priority for governmental, academic, and commercial research, with work in associated fields increasing rapidly ~\cite{refbib1,refbib2}. The National Quantum Initiative Act, signed into law in late 2018, provides a framework “to accelerate quantum research and development for the economic and national security of the United States” ~\cite{refbib2}, encouraging universities to accelerate their efforts to increase the nation's QISE capabilities through a grand unification of quantum computing algorithms ~\cite{refbib3} and deeper dives into quantum networking ~\cite{refbib4}. Meanwhile, commercial entities are focusing more and more on the development of fault-tolerant quantum computers ~\cite{refbib5} and quantum AI ~\cite{refbib6}.

The rise of viable quantum computers brings into question the sustainability of classical encryption methods. Most classical encryption methods rely on the mathematical complexity of large integer factorization to ensure security ~\cite{refbib7}; a complexity that may well become trivial with novel quantum phase estimation algorithms ~\cite{refbib8}.

To resolve this issue, quantum cryptography was first introduced by Stephen Wiesner in 1983 \cite{Wiesner}, and then it was popularized about a year later with the first specific QKD protocol ~\cite{refbib12}. The main advantage of QKD is that it encodes information in properties of the individual quanta used for transmission instead of distributing classical information through radio frequency (RF) or free-space optical (FSO) signal modulation  ~\cite{refbib9}. These quantum properties protect the encoded information from eavesdropper attacks via the Heisenberg Uncertainty Principle ~\cite{refbib10} and the No-Cloning Theorem ~\cite{refbib11}, both of which are foundational principles in quantum mechanics. While there are a number of potentially "quantum resistant" algorithms ~\cite{kyber} ~\cite{dili}, the fact that QKD attributes its security to fundamental physical principles instead of computational complexity allows the QKD-based encryption process to remain provably secure indefinitely, regardless of future increased computing capabilities.

There are two main categories of QKD: prepare-and-measure, and entanglement-based (quantum teleportation) ~\cite{pandm}. The former was the first proposed method of QKD, and although very practical and influential, it’s description is outside of the scope of this paper ~\cite{refbib12}. Within the realm of entanglement-based QKD, there have been a number of proposed protocols that use different algorithms to generate a secret encryption key. The most common of these protocols was developed by Artur Ekert in 1991, and it is aptly named E91 ~\cite{refbib13}.

\subsection{E91 Description} \label{e91}

The QKD model bit generation phase utilized in this work is based on the E91 Protocol introduced above. The original E91 Protocol was proposed for spin-$1/2$ particles such as electrons, but the logical parallel used for quantum optics utilizes polarization-entangled photons, therefore this is the variation modeled.

\begin{figure}[t]
	\includegraphics[width=0.48\textwidth]{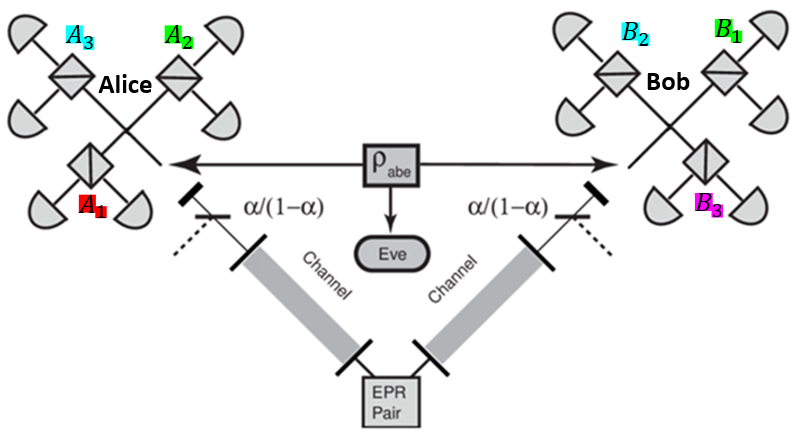}
	\centering
	\caption{Schematic of experimental setup for E91 (modified from \cite{zeevi}). Begins with entangled photon pair generation at the bottom, then photons are transmitted to Alice and Bob, who measure along one of three bases each.}
	\label{fig:fig1}
\end{figure}

The key distribution process begins with photon generation and transmission. An entangled photon pair is generated by an entanglement source, commonly through spontaneous parametric down-conversion (SPDC). A photon from an entangled pair is sent via a quantum channel to each of the two receivers, traditionally named Alice and Bob. (\cref{fig:fig1}).

\begin{figure}[ht]
	\includegraphics[width=0.48\textwidth]{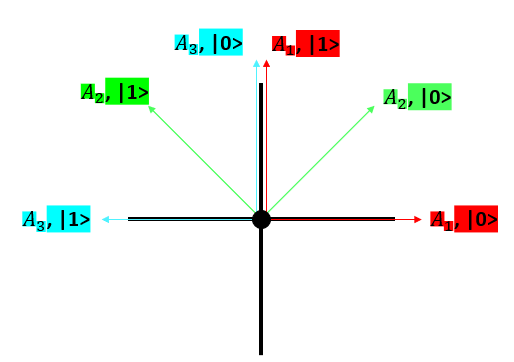}
	\centering
	\caption{Diagram of Alice's measurement bases for E91. The three bases are characterized by the  angles of their $\ket{0}$ states from horizontal.}
	\label{fig:fig2}
\end{figure}

\begin{figure}[t]
	\includegraphics[width=0.48\textwidth]{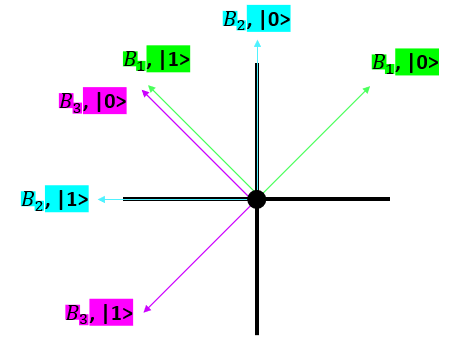}
	\centering
	\caption{Diagram of Bob's measurement bases for E91. The three bases are characterized by the  angles of their $\ket{0}$ states from horizontal.}
	\label{fig:fig3}
\end{figure}

Once Alice and Bob receive their photons, they randomly choose a polarization state basis to measure the received photon from a set of 3 bases each (denoted $A_{i}$ for Alice and $B_{j}$ for Bob, where ($i,j$) $\in \bigl\{1,2,3\bigl\}$). The different polarization bases are defined by the azimuthal angles of their $\ket{0}$ state from horizontal polarization: $A_{1}=0$, $A_{2}=\pi/4$, and $A_{3}=\pi/2$ for Alice (\cref{fig:fig2}), and $B_{1}=\pi/4$, $B_{2}=\pi/2$, and $B_{3}=3\pi/4$ for Bob (\cref{fig:fig3}). For each basis, measuring the polarization of the incident photon will result in either a $\ket{0}$ state or a $\ket{1}$ state, corresponding to a 0 or 1 bit, respectively.  Alice and Bob independently record their measured bit and the basis that they used for that measurement. This process is then repeated until a sufficiently large number of measurements have been made with respect to the desired final key length. Probabilistically, `sufficiently large' means at least $9/2$ times the desired length, as derived in \cref{sec:appen}

Once measurements are completed, Alice and Bob publicly share which bases they used for each measurement. They separate their measurements into two categories: one where they used the same bases (either $A_{2}$ and $B_{1}$ or $A_{3}$ and $B_{2}$), and one where they used different bases (all other combinations). For the same-basis measurements, the entanglement of the photons guarantees that Alice's and Bob's measurements are perfectly anti-correlated. Therefore, Bob need only invert his same-basis measurements to yield the same raw bitwise key that Alice achieved from her same-basis measurements.

Alice and Bob then use the different-basis measurements to determine if there was an eavesdropper, traditionally denoted as Eve, who intercepted their photons during transmission. They accomplish this using the generalized version of Bell's Theorem known as the Clauser-Horne-Shimony-Holt (CHSH) inequalities ~\cite{refbib14}.

To understand this process, we must first define the test statistic $S$ ~\cite{refbib13} ~\cite{refbib14} as
\begin{equation} \label{eq1}
S=E(a_{1},b_{1})-E(a_{1},b_{3})+E(a_{3},b_{1})+E(a_{3},b_{3}),
\end{equation}
where the correlation coefficient $E$ is ideally defined as
\begin{equation} \label{eq2}
E(a_{i},b_{j})=-a_{i}\cdot b_{j},
\end{equation}
but experimentally computed as
\begin{multline} \label{eq3}
E(a_{i},b_{j})=P_{1,1}(a_{i},b_{j})+P_{0,0}(a_{i},b_{j}) \\
-P_{1,0}(a_{i},b_{j})-P_{0,1}(a_{i},b_{j}).
\end{multline}

Here, $a_{i}$ and $b_{j}$ represent the unit vectors of the bases $A_{i}$ and $B_{j}$, respectively, and $P_{k,m}(a_{i},b_{j})$ represents the probability of measuring Alice's bit as $k$ and Bob's bit as $m$ given the measurement bases $A_{i}$ and $B_{j}$. Given the set of bases used by Alice and Bob, it can be quickly derived that the ideal $S$ value for perfect entanglement known as Tsirelson's Bound ~\cite{refbib15} is 
\begin{equation} \label{eq4}
S=-2\sqrt{2},
\end{equation}
while the ideal $S$ value for classical transmission is
\begin{equation} \label{eq5}
S=-2.
\end{equation}
For these full derivations, see \cref{sec:appen}

With these idealisms, we form the standard CHSH Bell Inequality as
\begin{equation} \label{eq6}
|S|\le 2,
\end{equation}
where a violation of this inequality points to a violation of local realism, indicating an acceptable entangled state.

When Eve measures the photon in a random basis, the state collapses accordingly. Then, when Eve attempts to retransmit this photon, there is now only a $1/3$ chance of choosing the right basis for transmission. This loss of information is the result of the No-Cloning Theorem ~\cite{refbib11}, and it means that the CHSH Bell inequality is no longer violated, indicating local realism.

Knowing this, Alice and Bob can then share their different-basis measurements and use them to compute the associated probabilities, correlation coefficients, and finally $S$ using \cref{eq1,eq3}. If \cref{eq6} is violated, then Alice and Bob determine that any discrepancy in S from the ideal value is due to noise, not Eve, meaning that their generated key is in fact secret. If, however, \cref{eq6} is true, then Alice and Bob determine that the quantum transmission was either too noisy, or local realism has been introduced to the transmission through Eve. In either case, they then discard all measurements, establish a new quantum channel, and repeat the measurement process to generate a new key.

This process describes an idealized situation where the photon source generates exactly one entangled photon pair per expected time bin, and where Alice and Bob receive and measure every single one of these photons in the same state as it was generated. This, of course, is not the case in real systems, where there are second photon pair generations, dark counts, loss, noise, and detector deficiencies. How the model deals with these practical shortcomings is discussed in \cref{sec:modelImplementation}\hspace{2.5mm}

E91 protocol only describes a method for bit generation, which is the first of three stages for the full QKD process. The subsequent stage known as information reconciliation corrects bit errors in the keys, and the final stage known as privacy amplification deals with the information leakage that results from error correction. The last two stages are not a focus of this paper, and thus, do not receive a full description here. However, the model's implementation of the two other stages is briefly described in \cref{sec:modelImplementation}\hspace{2.5mm} Additionally, potential improvements to these parts of the model are discussed in \cref{sec:conclusion}

\subsection{NASA Relevance}

Due to the significant complexity of NASA missions that are often expensive to build, run, and maintain, modeling serves an important role in the technology development process. The existing NQCAS software aims to mitigate some of these issues by modeling the processes NASA hopes to eventually perform in the lab. This informs NASA of what real-world conditions and parameters will lead to both expected and unexpected results. The QKD model that this paper describes is meant to validate  NQCAS entanglement distribution software through increased access to real-world experimental QKD data. % Introdução
\section{MODEL IMPLEMENTATION} \label{sec:modelImplementation}

\begin{figure*}[ht]
	\includegraphics[width=.8\textwidth]{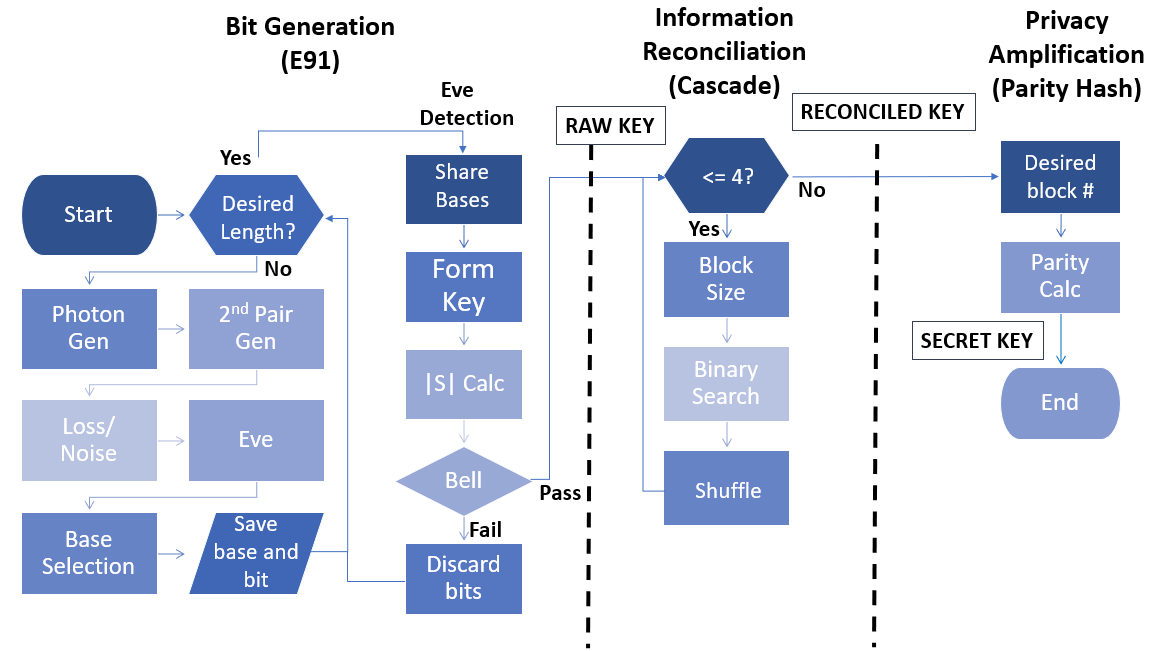}
	\centering
	\caption{Low-level abstraction for the NQCAS QKD model. Details the implementation of E91 for bit generation, Cascade for information reconciliation, and Parity Hash for privacy amplification.}
	\label{fig:fig4}
\end{figure*}

\begin{figure}[ht]
	\includegraphics[width=0.48\textwidth]{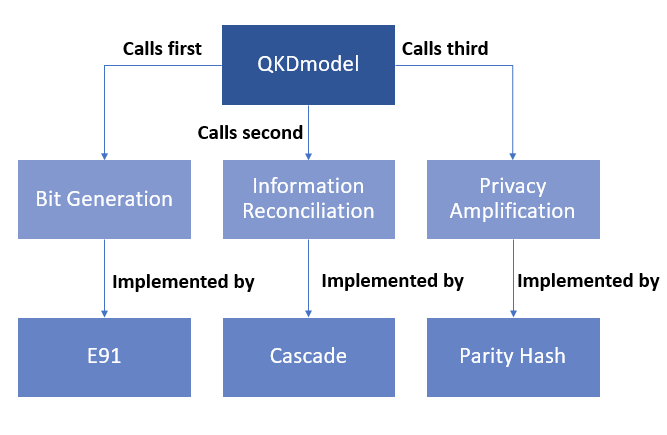}
	\centering
	\caption{High-level abstraction of the NQCAS QKD model. Illustrates the breakup of the model into three main stages that are then implemented by a specific protocol.}
	\label{fig:fig5}
\end{figure}

Section II is meant to provide an in-depth description of the QKD model's implementation by dividing the model into its three main stages: bit generation, information reconciliation, and privacy amplification (\cref{fig:fig4}).
Each stage has its own general function that implements the desired protocol for that stage and returns the associated statistics to the main `QKDmodel' runner (\cref{fig:fig5}).

The model has only one initial implementation for each stage, but in the future, the user will be able to select from many different implementations for each stage. This is further discussed in \cref{sec:conclusion}\hspace{2.5mm}

\subsection{Bit Generation}

To begin the process of bit generation, the first function called by `QKDmodel' (named `BitGeneration') takes only one input from the general runner: the desired key length. It then calls a specific implementation of bit generation to generate the associated statistics. This function outputs Alice's and Bob's raw keys, the raw key rate (coincidences per second), the raw quantum bit error rate (QBER) (correct measurements / incorrect measurements), the calculated S value, and the simulated elapsed time to return to the `QKDmodel' function. The initial specific implementation of `BitGeneration' is a simulation of E91.

The E91 simulation begins by defining various relevant constants. These constants include Alice's and Bob's sets of possible bases, the appropriate Tsirelson Bound (calculated from the basis unit vectors using \cref{eq1,eq2}), and the excess bit factor (conventionally 2), as well as device-and-path specific parameters such as SPDC pump rate, photon generation rates, and relevant loss and noise values.

Once the relevant constants are defined, the model uses MATLAB's built-in algorithms to generate pseudorandom numbers to arbitrarily choose a basis for Alice to measure in from the constant basis set. It also randomly chooses the qubit that Alice measures as a 0 or a 1. Hence, the quantum state is modeled by two characteristics: a single bit, and the basis that Alice measures it in.

	After the Alice basis and bit generation, the measurement of the state is modeled using either one or two Bell State collapses, dependent on whether or not that transmission has been intercepted by Eve (a constant probability per pump). The state collapse process is informed by the E91 and CHSH papers \cite{refbib13,refbib14}, where they define \cref{eq1,eq2,eq3}. The scalar product of Alice's and Bob's basis unit vectors are used to compute the theoretical same-bit probability. Subsequently, the bit measurements are determined for whoever is making said measurement. For the same-bit probability derivation, see \cref{sec:appen}
 
 In the Eve case, the collapse is first performed using a random Bob basis as the receiver and Alice's state as that which will be used to model Eve's measurement. Next, the measurement is performed again using another randomly generated Bob basis to determine Bob's bit measurement. This time, however, the incoming state is modeled by the outputted state of the first collapse (Eve's collapse) instead of Alice's, which removes the entanglement correlation.
 
 In the non-Eve case, Bob's measurement is modeled simply with one collapse instead of two, using Alice's bit and basis as the incoming state as intended. In either case, the theoretical maximum S value for the transmission is used to enforce the appropriate Tsirelson bound (\cref{eq4}).
 
	Once the collapse of the transmitted state is complete with the Eve implementation, the next step is to model the non-Eve loss and noise. This is done using loss and noise input parameters from NQCAS. These parameters consist of the effective detection efficiency $\eta_{d}$, the noise count probability $v_{d}$, and a photon number resolution (PNR) parameter $\rho_{d}$, which is 0 for no PNR and 1 for ideal PNR ~\cite{refbib16}.
 
 These parameters are then used to calculate a positive operator-value measure (POVM), which informs the photon reception probabilities:
 
 \begin{equation} \label{eq7}
P(0|n)=(1-v_{d})[1-\eta_{d}]^n
\end{equation}
 \begin{multline} \label{eq8}
P(1|n)=v_{d}[1-\rho \eta_{d}]^n+ \\
(1-v_{d}) \displaystyle\sum_{k=0} ^{n-1} \eta_{d}(1-\eta_{d})^k(1-\rho_{d} \eta_{d})^{n-1-k}.
\end{multline}
 Here, $P(k|n)$ is the probability of the detector registering $k$ photons given $n$ incident source photons.
 
 If zero detections are registered, then the received bit becomes null (indicating loss). Otherwise, $v_{d}$ is used to inform the probability of a bit flip due to noise. If the bit is not null or flipped due to noise, then the modeled qubit has successfully been transmitted. Once again, the calculated Tsirelson bound is used to cap the $S$ value for low-bit or high-noise transmission.
 
	After the noise and loss implementation is complete, the next step is to separate the same-basis measurements from the different-basis measurements. All coincident same-basis measurements are saved and used to generate the raw key, and all coincident different-basis measurements are saved separately for later use.
 
	This whole process repeats until the desired key length multiplied by the excess bit factor bits are achieved in the same-basis measurements. Then, the simulated elapsed time is calculated as entangled photon pairs / pump rate. Finally, the different basis measurements are used in the CHSH inequality test to calculate S, as described in \cref{sec:introduction}\hspace{2.5mm} The raw data from the simulation is then returned to `BitGeneration', where the desired statistics are calculated and subsequently returned to the `QKDmodel' function.

\subsection{Information Reconciliation}

The second function called by `QKDmodel' (named `InformationReconciliation') takes as its input each of the raw keys, the desired number of reconciling iterations, the number of seconds it took to generate the raw keys, and the raw QBER. The function returns Alice's and Bob's reconciled keys, the reconciled key rate (uncompromised bits / second), the reconciled corrected QBER (corrected bits / total bits), and the reconciled uncorrected QBER (remaining errors / total bits) to the `QKDmodel' function. The initial implementation of `InformationReconciliation' is a simulation of the Cascade Protocol called `Cascade' ~\cite{refbib17}.
	
		The protocol begins by calculating the starting block size as

 \begin{equation} \label{eq9}
blockSize(1,QBER)=0.73/QBER.
\end{equation}
Making the block size smaller than $1/QBER$ in this way makes it more likely that there will be no more than one error per block, which is advantageous because Cascade cannot correct blocks that have more than one error.

For each subsequent iteration, however, the block size is calculated as
 \begin{equation} \label{eq10}
blockSize(k, QBER)=2*blockSize(k-1,QBER),
\end{equation}
where k is the current iteration number. This is because each iteration of Cascade corrects some number of errors, lowering the QBER and allowing for a larger block size ~\cite{refbib17}.

 The model then steps into a separate function to perform a recursive binary search of Bob's key, using the even parity of each block as its error check as compared to Alice's. When an error in parity is found, it divides the block size in two and then repeats the algorithm recursively until the basis case of a 2 bit block. It corrects the incorrect bit, thus compromising 2 bits in the key for each correction. This compromised information is known as information leakage, and it will be discussed further in the next subsection.
  
	Each recursive search returns the reconciled keys and information leakage for that search to `Cascade'. Alice's and Bob's keys are then shuffled in the same order using a randomly generated MATLAB shuffle algorithm, and the whole process repeats the desired number of times (usually 4).
 
 The reconciliation data is then returned to `InformationReconciliation', which uses the leakage, the remaining errors, and the time duration of the key generation to calculate the statistics described above. These statistics are then returned to `QKDmodel'.

\subsection{Privacy Amplification}

The final function called by `QKDmodel' (named `PrivacyAmplification') takes as its input each of the reconciled keys, the desired key length, and the number of seconds it took to generate the raw keys, and it outputs Alice's and Bob's final secret keys and final secret key rate (secret bits/second). The initial implementation of `PrivacyAmplification' is a simple original protocol named `ParityHash'.

		The protocol begins by splitting the key into blocks such that there are exactly as many blocks as the desired key length. To do this, the model calculates the block size as
  
 \begin{equation} \label{eq11}
 blockSize=\lfloor n/d \rfloor,
\end{equation}
where n is the length of the reconciled keys and d is the desired key length. For the number of blocks to match the desired length, the length of the final block is the number of block-less bits remaining after the initial calculation.
  
  The model then populates the final secret key with the even parity of each block, leaving Alice and Bob's secret keys at the desired length. These are then returned to `PrivacyAmplification', where the associated statistics are calculated and returned to `QKDmodel'.
 % Estrutura do Trabalho
\section{RESULTS AND ANALYSIS} \label{sec:results}

Section III displays the preliminary results of the simulation, and analyzes these results with respect to similar data collected by Ling ~\cite{ling}, Fujiwara ~\cite{fuji}, and Marcikic ~\cite{marc}. The objective of these preliminary simulations is to compare the simulated data to the published data to assess the viability of the simulation. Therefore, the loss and noise input parameters used in the model for these simulations are loosely informed by the relevant literature ~\cite{ling,fuji,marc}.

 The non-Eve simulations are generated using the following input parameters: a key length between 295-300 bits, a pump rate of 1 GHz, a first pair generation rate of $4*10^{-6}$ per pump, and a high second pair generation rate of 1/3 per first pair, as well as a noise count probability of $2*10^{-4}$, a detector efficiency of 80\%, and a PNR parameter $\rho_{d}$ of 0.8. The Eve statistics are generated using the same parameters, except with key lengths between 45-50 bits to reduce computation time (see \cref{sec:conclusion}\hspace{2.5mm}). These parameters are used to generate three main statistics: S value, QBER, and key rate.

\subsection{S Value}

The first statistic returned by the model is the S value. This is an indication of how well the quantum state survived the transmission/reception, and is typically used to detect Eve.

\begin{figure}[ht]
	\includegraphics[width=0.48\textwidth]{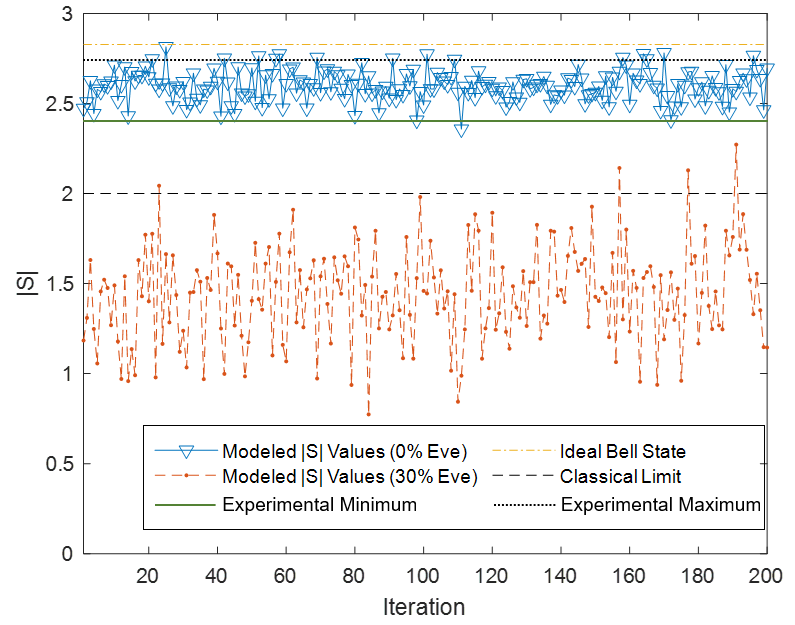}
	\centering
	\caption{Modeled $|S|$ values from the Monte Carlo simulation in comparison with the ideal bounds. $X\%$ Eve indicates an $X\%$ chance of an eavesdropper intercepting any given photon.}
	\label{fig:fig6}
\end{figure}

The modeled S value data is shown in \cref{fig:fig6}. For the non-eavesdropper (0\% Eve) scenario, the model yields results with a mean that is about 4\% off of the experimental means of Ling ~\cite{ling} and Fujiwara ~\cite{fuji}. This indicates that the current quantum state collapse implementation is reasonable for the non-Eve case.

Additionally, when an eavesdropper is introduced to the model with a 30\% chance of intercepting any given photon (30\% Eve), we can see the modeled S value plummets below the classical limit, indicating the introduction of local realism. This indicates that the current quantum state collapse implementation is reasonable for the Eve case as well.

The 4 data points for this scenario that violate the classical limit indicate scenarios when Alice and Bob would attribute their measurement disruptions to high noise and not Eve. These scenarios become increasingly unlikely as the key length increases, which will be needed for complete NQCAS verification testing. Additionally, the high second pair generation rate of 1/3 per first pair does not degrade the link as much as one may expect. These topics are addressed in \cref{sec:conclusion}

\subsection{QBER}

The second statistic returned by the model is the raw QBER. This is a measurement of the ratio of incorrect measurements to correct ones, and therefore, it is not a statistic that would be able to be measured for real-world E91 implementation without compromising security. However, it is a useful figure to know for experimentation when researchers are more concerned with establishing a quantum link and producing the associated link budget than actual secure communication.

\begin{figure}[ht]
	\includegraphics[width=0.48\textwidth]{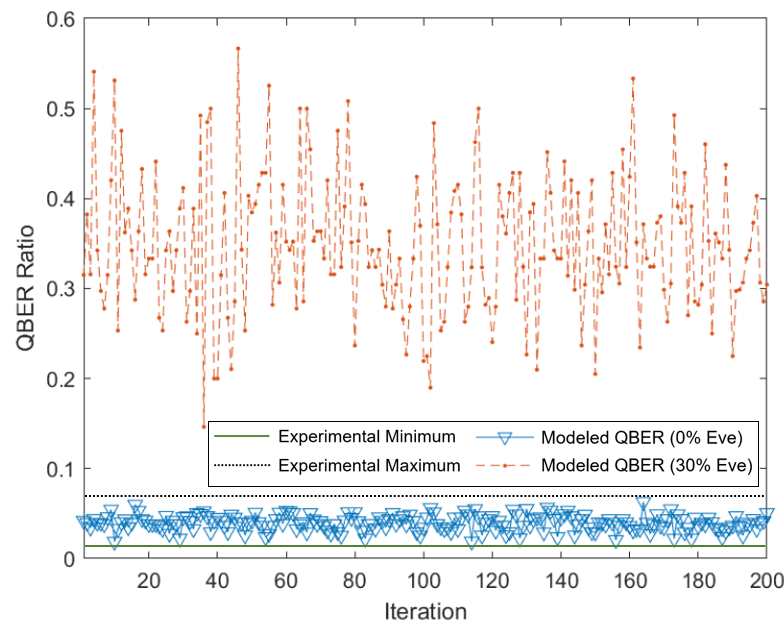}
	\centering
	\caption{Modeled raw QBER values from the Monte Carlo simulation. $X\%$ Eve indicates an $X\%$ chance of an eavesdropper intercepting any given photon.}
	\label{fig:fig9}
\end{figure}

The modeled raw QBER data is shown in \cref{fig:fig9}. For the non-eavesdropper scenario, the model yields results with a mean that is about 3\% off of the experimental means of Ling ~\cite{ling}, Fujiwara ~\cite{fuji}, and Marcikic ~\cite{marc}. This indicates that the current loss and noise implementation is reasonable for the non-Eve case. Also, as with the S value data, the 30\% Eve case is also simulated. For this scenario, we see the modeled QBER jump, as expected, to between 20\% and 50\%, indicating larger amounts of error. Since the receivers are unable to distinguish between Eve noise and background noise, this indicates that the current loss/noise implementation is also reasonable for the Eve case.

\subsection{Key Rates}

The final statistics returned by the model are the key rates. The raw key rate is a measure of how many coincidences occurred per second (from bit generation), the reconciled key rate describes how many uncompromised bits occur per second (from information reconciliation), and the secret key rate describes how many secret key bits are generated per second (from privacy amplification).

\begin{figure}[t]
	\includegraphics[width=0.48\textwidth]{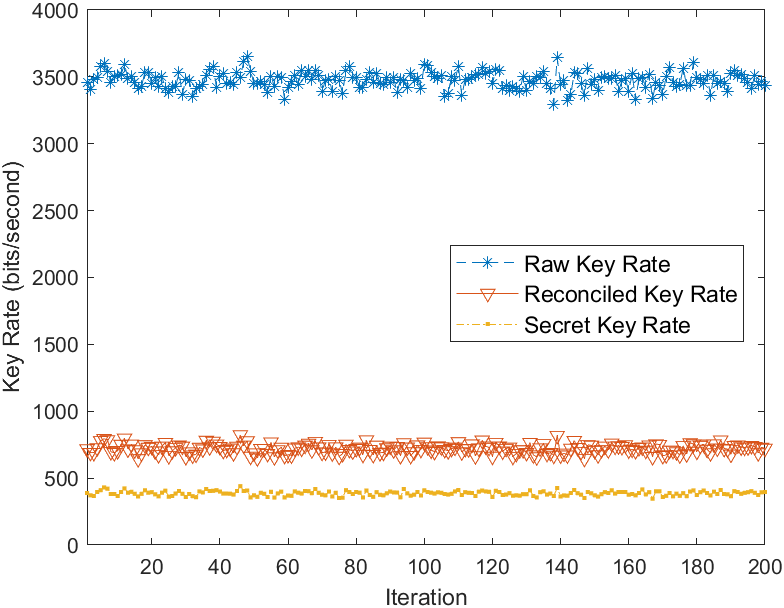}
	\centering
	\caption{Modeled key rates from the Monte Carlo simulation. The raw, reconciled, and secret key rates are products of the bit generation, information reconciliation, and privacy amplification stages respectively.}
	\label{fig:fig13}
\end{figure}

The simulated key rates are shown in \cref{fig:fig13}. All three key rates yield results on the same order of magnitude as the expected values from Ling \cite{ling} and Marcikic \cite{marc}. However, high variability in the experimental data makes it difficult to define bounds and averages, which makes it difficult to compare the data. As a result, there are a few notable discrepancies between the modeled data and the experimental data for the key rates. First, the modeled reconciled key rate is slightly lower than the Marcikic's data would suggest, and this is because Marcikic uses a more efficient variation of the Cascade protocol than is implemented in the model. Additionally, the data from the literature show random fluctuations in key rate at various times that can be attributed to environmental factors like the sunrise, which is not modeled. Finally, the experimental data are generated using continuous wave (CW) lasers, while the model is based on the use of a pumped laser, which makes it difficult to translate between the two perfectly. Each of these differences will be addressed in \cref{sec:conclusion}

 %Outras Normas
\section{FUTURE STEPS AND CONCLUSION} \label{sec:conclusion}

Section IV concludes with a summary of the work described in this paper, as well as where this effort will progress in the future.

This paper described a QKD model that will eventually be used to verify the NQCAS software for entanglement distribution and entanglement swaps. It provided an introduction to QKD through the explanation of the E91 protocol, and described in detail the implementation of a QKD model. It also provided some preliminary data collected from said model, and analyzed the modeled data in comparison to published experimental QKD data.

In the future, there are a number of different steps that may be taken to increase the utility of the work described. First, adjusting the implementation of the model such that it considers different types of source lasers and different environmental noise sources, allowing the model to output data that more closely reflects the experimental data. Next, vectorizing the MATLAB scripts to optimize efficiency, as opposed to the current iterative method, to reduce runtime. Additionally, running more simulations with varying input parameters to provide insight into discrepancies between the simulation and what was expected, specifically with the simulated $|S|$ values and the second photon pair generation. Finally, implementing additional protocols at each of the three stages, allowing comparison to more published data for better verification. For bit generation, another common entanglement-based QKD protocol in the literature is known as the BBM92 protocol ~\cite{refbib21}. There are also prepare-and-measure protocols that have been modified to become entanglement-based, like the SARG04 protocol ~\cite{refbib22,refbib23}. For information reconciliation, implementing a variation of Cascade similar to that in Marcikic's work \cite{marc} would allow the reconciled key rate to match the published data more closely. Additionally, there are other information reconciliation protocols used in the relevant literature such as turbo codes and low-density parity checks ~\cite{refbib24, refbib25}. For privacy amplification, there are a variety of hashing functions used in experiments to make up for the leakage from the reconciliation stage ~\cite{refbib26}. For example, many experiments use need-based hashing to find a pertinent hash function based on the amount of leakage relative to the key size, which would also be useful to implement ~\cite{refbib27}.

This work is a step towards the verification of an entanglement simulation that will be used to inform a wide range of quantum networking experiments preformed at NASA Glenn Research Center. These results will help to support the overarching NASA mission, and they will help to push the boundaries of science in the years to come. % Agradecimentos
\section{APPENDIX} \label{sec:appen}

Section V is meant to provide some of the specific derivations referenced in the paper. It begins with a calculation of the probabilistic excess measurement factor with respect to the desired key length, then it provides the calculation of Tsirelson's bound and the classical limit using the Ekert bases, and finally, it provides the calculation for same-bit measurements that was used to inform the modeling of the quantum state collapse.

\subsection{Excess Measurement Factor}

Let $P_{same}$ be the probability of Alice and Bob choosing the same basis for any give measurement, let $L$ be the desired key length, and let $N$ be the total number of measurements. We then form the equation

\begin{center}
$L=P_{same}N$.
\end{center}

From here, $P_{same}$ can be broken down into the probability of Alice choosing $a_{2}$ and Bob choosing $b_{1}$ ($P_{a_{2},b_{1}}$) plus the probability of Alice choosing $a_{3}$ and Bob choosing $b_{2}$ ($P_{a_{3},b_{2}}$):

\begin{center}
$P_{same}=P_{a_{2},b_{1}}+P_{a_{3},b_{2}}$.
\end{center}

Similarly, $N$ can be broken down into the excess measurement factor $K$ times the desired key length:

\begin{center}
$N=KL$.
\end{center}

Combining and isolating yields

\begin{center}
$L=(P_{a_{2},b_{1}}+P_{a_{3},b_{2}})KL$
\end{center}

\begin{center}
$K=\frac{1}{P_{a_{2},b_{1}}+P_{a_{3},b_{2}}}$.
\end{center}

Knowing that Alice and Bob each choose 1 of 3 bases for the Ekert protocol, this yields the final result for the excess measurement factor $K$:

\begin{center}
$K=\frac{1}{\frac{1}{3}\frac{1}{3}+\frac{1}{3}\frac{1}{3}}$
\end{center}

\begin{center}
$K=\frac{9}{2}$.
\end{center}

\subsection{Theoretical S Values}

Recall \cref{eq1,eq2} from \cref{sec:introduction}\hspace{2.5mm} Plugging \cref{eq2} into \cref{eq1} yields a theoretical S value such that

\begin{center}
$S=-a_{1}\cdot b_{1}+a_{1}\cdot b_{3}-a_{3}\cdot b_{1}-a_{3}\cdot b_{3}$.
\end{center}

For entanglement distribution, $a_{i}$ and $b_{j}$ are the basis unit vectors as described in \cref{sec:introduction}\hspace{2.5mm} Therefore, Tsirelson's bound can be calculated from the dot products of the appropriate unit vectors:

\begin{center}
$S=-\begin{bmatrix} 1 & 0 \end{bmatrix}\cdot \begin{bmatrix} \frac{1}{\sqrt{2}} \\ \frac{1}{\sqrt{2}} \end{bmatrix}+\begin{bmatrix} 1 & 0 \end{bmatrix}\cdot \begin{bmatrix} \frac{-1}{\sqrt{2}} \\ \frac{1}{\sqrt{2}} \end{bmatrix}$
\end{center}
\begin{center}
$-\begin{bmatrix} 0 & 1 \end{bmatrix}\cdot \begin{bmatrix} \frac{1}{\sqrt{2}} \\ \frac{1}{\sqrt{2}} \end{bmatrix}-\begin{bmatrix} 0 & 1 \end{bmatrix}\cdot \begin{bmatrix} \frac{-1}{\sqrt{2}} \\ \frac{1}{\sqrt{2}} \end{bmatrix}$
\end{center}
\begin{center}
$S=-2\sqrt{2}$.
\end{center}

For classical distribution, the lack of superposition negates the need for unit vectors, as all of the measurement bases can be viewed as the same, resulting in dot products of 1. Plugging in 1 for each dot product yields the classical $S$ value limit:

\begin{center}

$S=-1+1-1-1$\\
$S=-2$.

\end{center}

\subsection{Same-Bit Probability}

Recall \cref{eq3} from \cref{sec:introduction}\hspace{2.5mm} The probabilities in this equation can be split into 2 categories: one where Alice and Bob have the same bit measurement, and one where they do not. For any given basis combination, let $P_{=}$ be the sum of the former category and let $P_{x}$ be the sum of the latter such that
\begin{center}
$P_{=}=P_{0,0}+P_{1,1}$\\
$P_{x}=P_{1,0}+P_{0,1}$.
\end{center}

According to \cref{eq2,eq3}, this means that the theoretical correlation coefficient $E$, in terms of the theoretical collapse probabilities, is given by
\begin{center}
$E(a_i,b_j)=-a_i\cdot b_j=P_{=}-P_{x}$.
\end{center}

When this equation is combined with the normalization condition of
\begin{center}
$P_{=}+P_{x}=1$,
\end{center}
it can then be used to solve for the ideal probability for same-bit measure in terms of the theoretical correlation coefficient:
\begin{center}
$1+E(a_i,b_j)=2P_=$.
\end{center}

Therefore, the final same-bit probability as it is implemented in the model can be written as
\begin{center}
$P_==\frac{1-a_i\cdot b_j}{2}$
\end{center}
for the measurement bases $A_i$ and $B_j$.
\newline
\newline
\begin{acknowledgments}

I would like to thank both the Space Communications and Navigation Internship Program and everyone at NASA Glenn Research Center who has helped me with this research. Especially: Ian Nemitz, Yousef Chahine, Evan Katz, Brian Vyhnalek, Adam Fallon, John Lekki, Bryan Welch, Chris Nadeau, Ian Chin, Dilara Sen, Tim Gallagher, and Bart Upah. Thank you for making my internship experience both engaging and rewarding.

All original figures were made using Microsoft PowerPoint and Microsoft Paint. All code was run, and all original plots were generated using MATLAB R2023a.

\end{acknowledgments}
\bibliographystyle{setup/bib_sophia}
\bibliography{references}
 %Referências

\end{document}